\documentclass[aps,prl,twocolumn,amsmath,amssymb,floatfix]{revtex4}
\usepackage{dcolumn}
\usepackage{bm}
\usepackage{graphicx}
\usepackage{epstopdf}
\usepackage{color}

\def\vp{{\mbox{\boldmath$p$}}}
\def\vs{{\mbox{\boldmath$s$}}}
\def\vB{{\mbox{\boldmath$B$}}}
\def\vR{{\mbox{\boldmath$R$}}}

\def\tDelta{{\tilde\Delta}}
\def\pfhv{{\mbox{\boldmath$p$}}_{\!{\scriptscriptstyle F}}}

\def\vvF{{\mbox{\boldmath$v$}}_{\!{\scriptscriptstyle F}}}

\begin{document}

\renewcommand{\section}[1]{{\par\it #1.---}\ignorespaces}

\title{Spontaneously broken time-reversal symmetry in high-temperature superconductors}

\author{Mikael H{\aa}kansson}
\author{Tomas L\"ofwander}
\author{Mikael Fogelstr\"om}
\affiliation{Department of Microtechnology and Nanoscience - MC2,
Chalmers University of Technology, SE-412 96 G\"oteborg, Sweden}
\date{\today}

\begin{abstract}
Conventional superconductors are strong diamagnets that through the Meissner effect expel magnetic fields.
It would therefore be surprising if a superconducting ground state would support spontaneous magnetics fields.
Such time-reversal symmetry broken states have been proposed for the high-temperature superconductors, but
their identification remains experimentally controversial.
Here we show a route to a low-temperature superconducting state with broken time-reversal symmetry
that may accommodate currently conflicting experiments. This state is characterised by an unusual vortex
pattern in the form of a necklace of fractional vortices around the perimeter of the material,
where neighbouring vortices have opposite current circulation.
This vortex pattern is a result of a spectral rearrangement of current carrying states near the surfaces.
\end{abstract}

\maketitle

The phase-sensitive experiments \cite{Wollman1993,Tsuei1994} carried out in the early 1990's showed
that the high-temperature superconductors have predominantly $d$-wave pairing symmetry.
Ever since, there has been an ongoing debate whether there exists a low-temperature phase that
breaks time-reversal ($\cal{T}$) symmetry, in addition to the reflection symmetry of the crystal broken by the $d$-wave
order parameter itself.
Several experiments on tunnelling and charge transport support a phase transition into a state with broken ${\cal T}$-symmetry 
at low temperatures \cite{Covington1997, Krishana1997, Dagan2001, Gonnelli2001, Elhalel2007, Gustafsson2013}.
On the other hand efforts to detect the concomitant spontaneous magnetic field have failed,
or at best have put severe restrictions on how strong the subdominant order
can be \cite{Tsuei2000, Carmi2000, Neils2002, Kirtley2006, Saadaoui2011}. 
Below we show how to reconcile these two sets of experiments and how the route
to a low-temperature superconducting state with broken ${\cal T}$-symmetry may occur. 
We emphasize that the superconducting state with broken ${\cal T}$-symmetry we describe here
does not necessarily imply a multicomponent superconducting order-parameter \cite{Yip1993,Schemm2014}.

\begin{figure*}[tbh]
\includegraphics[width=\columnwidth]{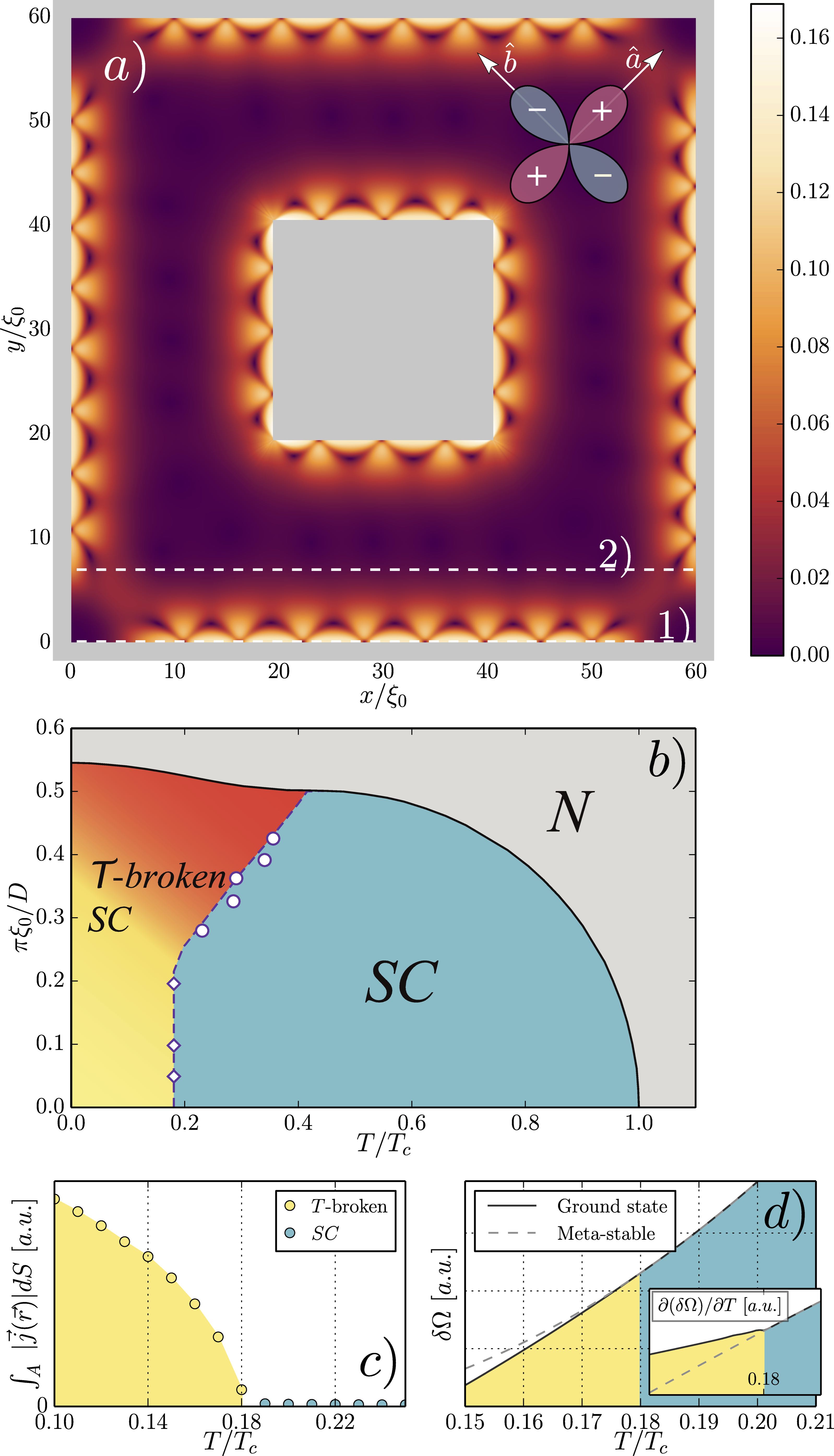}
\caption{
(a) A square grain of a $d$-wave superconductor including a hole in the center.
The crystal $ab$-axes are rotated by 45$^\circ$ relative to the grain surfaces.
At low temperatures (here $T=0.1T_c$) ${\cal T}$-symmetry is spontaneously broken as manifested by
fractional vortices near the surfaces. The color scale shows the magnitude of the circulating currents in units of the depairing 
current $j_d=2\pi e N_F v_F k_B T_c$.
(b) Phase diagram for $d$-wave superconducting grains as function of temperature and inverse grain size (side length $D$). 
We find that the low-temperature phase breaks ${\cal T}$-symmetry. The circles mark a transition into a state that only 
breaks ${\cal T}$-symmetry. For large $D$ the state also breaks translational symmetry along the surface
and the transition is marked by diamonds. The transition between the two states is not distinct as indicated by the color gradient.
The transition (solid line) between the normal state (N) and the superconducting states (SC) is taken from Ref.~\onlinecite{anton}.
(c) Temperature dependence of the magnitude of the currents integrated over the grain area. For large
grains, here $D=60\xi_0$, the transition temperature $T_{cs}\approx 0.18 T_c$ is rather sharp and independent of $D$.
(d) The free energy of the grain as function of temperature.
The dashed curve corresponds to the meta-stable state with conserved ${\cal T}$-symmetry.}
\label{fig:phasediagram}
\end{figure*}

The superconducting state in unconventional superconductors, such as the cuprates, is fragile to scattering
off impurities, defects and surfaces \cite{Buchholtz81}.
Scattering leads to pair-breaking and formation of so-called Andreev states \cite{Buchholtz81,Hu1994,Lofwander2001}.
These states are formed by constructive normal and Andreev reflection processes,
have energies $\varepsilon_A$ within the superconducting energy gap $\Delta$, and are spatially bound to the scattering centers.
For superconducting grains, when the size of the superconducting material is comparable
with the superconducting coherence length, the whole superconducting state of the grain is affected by boundary scattering
and the formation of Andreev states leads to properties that are not fully understood yet.
For the realization of real devices, a deeper understanding of the ground state of grains is therefore called for.

\begin{figure}[tbh]
\includegraphics[width=\columnwidth]{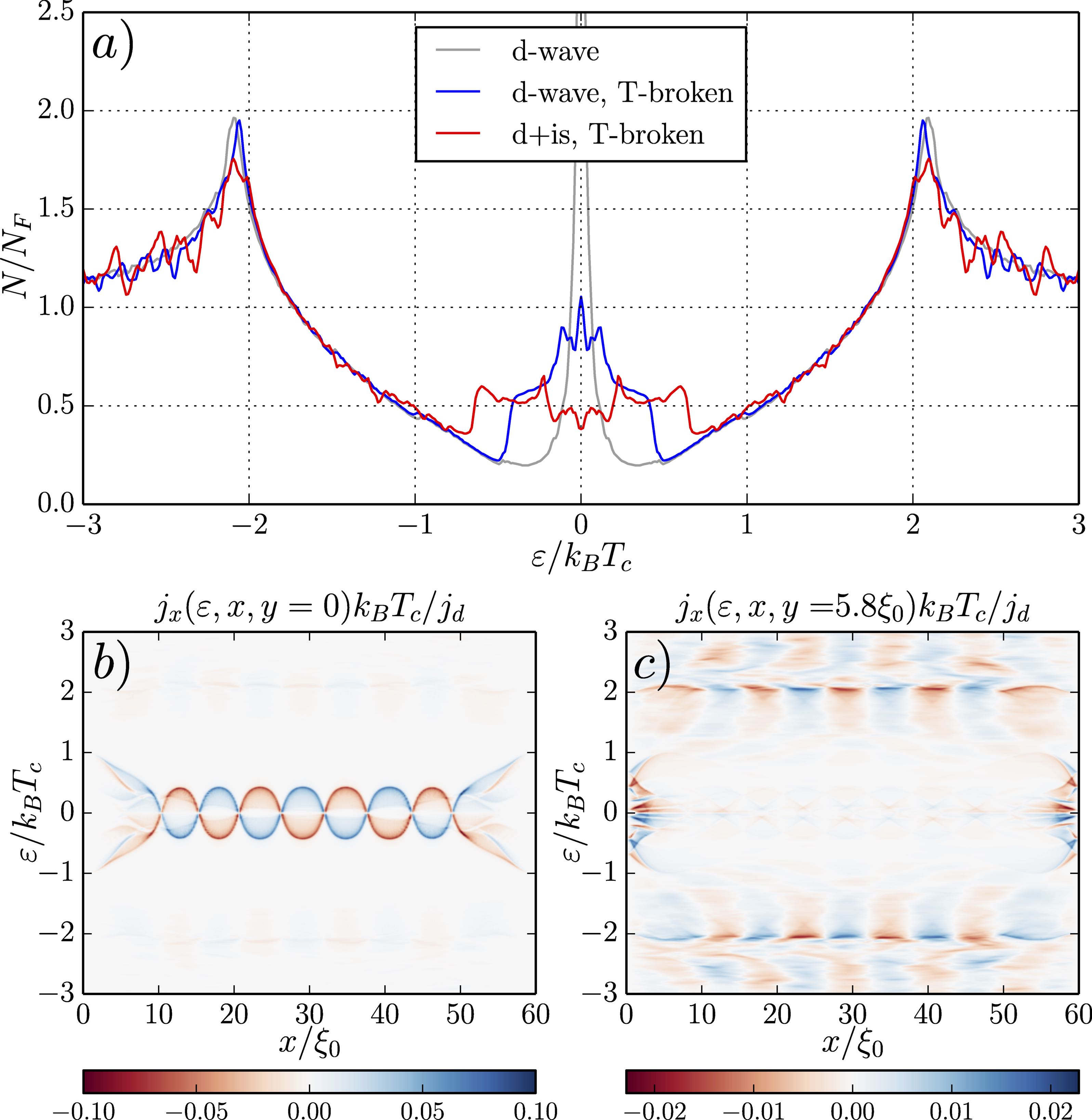}
\caption{\label{fig:DOS} 
(a) The density of states integrated over the grain area at temperatures $T=0.25T_c$  (grey line)
and $T=0.1T_c$ (blue line). Both are for a pure $d$-wave order parameter. For comparison we also show the 
case of a d+is phase (red line). In this case the broad peak around $\varepsilon=0$ splits into two.
(b)-(c) The spectral currents in the two cross-sections of the grain shown by dashed white lines in Fig.~\ref{fig:phasediagram}(a).
The currents near the surface (b) is carried by the subgap part of the spectrum (Andreev surface states), while further
away from the surface at $y=5.8\xi_0$ (c), the currents are carried by continuum states.}
\end{figure}

\begin{figure*}[tbh]
\includegraphics[width=\columnwidth]{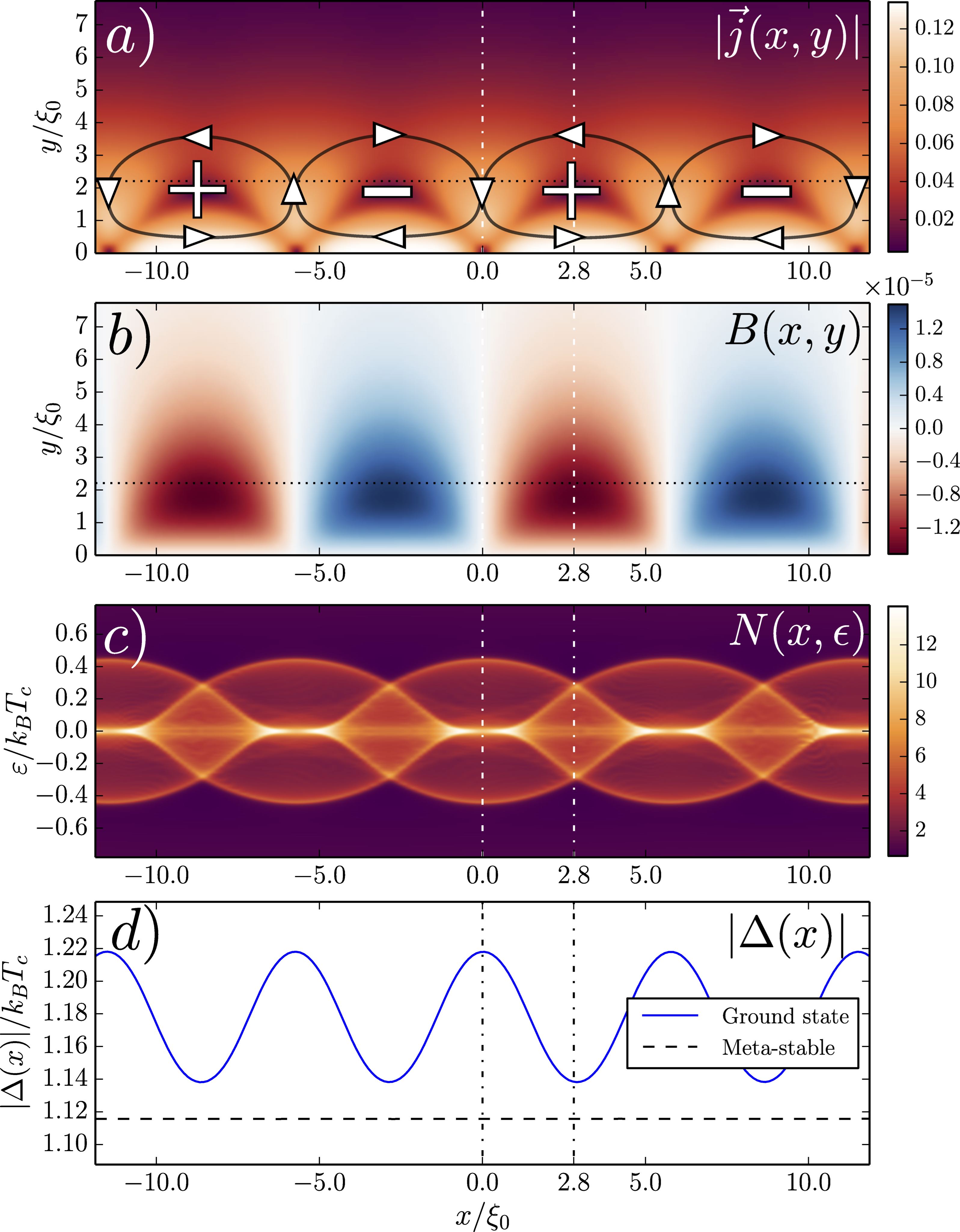}
\caption{\label{fig:sprops} 
(a) Details of the circulating currents, showing the staggered order.
(b) The circulating currents give rise to a magnetic field pattern forming fractional vortices containing
fluxes between $\sim\pm 10^{-5}\Phi_0$. 
(c)-(d) The spatial variation of the spectrum of low-lying states and the magnitude of the order parameter at $y\approx 2 \xi_0$
away from the surface along the black dotted line in (a). The units are $j_d$ in (a), $\Phi_0/\xi_0^2$ in (b), and $N_F$ in (c).}
\end{figure*}

We consider meso-scaled grains of a $d$-wave superconductor and relax the assumption of translational invariance along
the surface. The typical grain sizes we consider correspond to side lengths $D\sim10\xi_0$ to $100\xi_0$,
where $\xi_0=\hbar v_F/(2\pi k_B T_c)$ is the superconducting coherence length ($v_F$ is the quasiparticle velocity
at the Fermi surface in the normal state).
We focus on rotated crystals with 45$^\circ$ misaligned surface-to-crystal orientation, such that the $d$-wave order
parameter nodes are pointing towards the surfaces and pair-breaking effects are most pronounced. 
In this case the order parameter is reduced to zero along the surface of the grain. We study grains with holes,
as shown in Fig.~\ref{fig:phasediagram}(a), and also grains without holes. 

At low temperature, ${\cal T}$-symmetry and translational symmetry along the surface are spontaneously broken.
In this state the order parameter acquires local phase gradients and a necklace-like vortex pattern of currents appears near the surface.
The fractional vortices along the surfaces of the grain are ordered in a staggered fashion, such that
neighbouring vortices have opposite current circulation.
This inhomogeneous phase has lower free energy than the thin-film phase in Ref.~\onlinecite{anton} and has
a high critical temperature $T_{cs}$.

The transition is found for all sizes of grains we have considered (up to $600\times 600\, \xi_0^2$), with or without holes.
In Fig.~\ref{fig:phasediagram}(c) we show the magnitude of the spontaneous currents
integrated over the grain area as function of temperature.
The temperature of the transition below which the currents appear is around $0.18 T_c$ independent of grain size
as long as $D>\xi_0$, see the phase diagram in Fig.~\ref{fig:phasediagram}(b). 
This state is stable for surface-to-crystal mis-orientations from $45^\circ$ down to $\sim 23^\circ$.
The temperature dependence of the free energy (measured relative to the free energy of the non-superconducting state)
for a grain of size $D\approx 60\xi_0$ is displayed in Fig.~\ref{fig:phasediagram}(d).
Its temperature derivative (entropy difference between normal and superconducting states) is shown as inset.
The vortex phase is energetically favourable and the phase transition is of second order.

At high temperatures, the grain area-integrated density of states (DOS)
has a large peak at the Fermi level, see grey curve in Fig.~\ref{fig:DOS}(a).
The peak is due to the well known zero-energy states formed by Andreev scattering off order parameter lobes
of different signs \cite{Hu1994}.
In the symmetry broken state the low-energy states are rearranged
so that the spectral weight is shifted symmetrically away from the Fermi level, see blue curve in Fig.~\ref{fig:DOS}(a).
This shifting of spectral weight away from the Fermi level is due to the spontaneous
currents appearing in an area close to the surface. These currents are driven by a finite local phase gradient
of the order parameter, which leads to Doppler shifts $\vp_s(\vR)\cdot \vvF$ with $ \vp_s(\vR)=\frac{\hbar}{2} \nabla \phi(\vR)$.
The Doppler shifts are the source of the lowering of the free energy in the ${\cal T}$-symmetry broken state.
The transition temperature $T_{cs}$ is only limited by the maximum Doppler shifts the superconductor can sustain,
which corresponds to phase gradients and currents of the order of the depairing current
near the surfaces where the order parameter is suppressed.

The spectral rearrangements in the vortex phase also leads to interesting rearrangements of the spectral current density,
as shown in Fig.~\ref{fig:DOS}(b)-(c). The currents near the surfaces are carried by the subgap part of the spectrum
and the direction along the surface alternates, as seen by the meandering of the quasiparticle states along the interface.
The color scale includes both positive currents ($j_x>0$, red color) and negative currents ($j_x<0$, blue color).
In panel \ref{fig:DOS}(c) we show the local spectral-current density at a distance $\sim 6\xi_0$ from the surface.
The currents inside the grain are carried by continuum states close to the gap energy.
These currents flow in the opposite direction compared to current right at the surface and they are carried by the condensate.
Together, these spectral currents lead to the circulating current flow pattern displayed in Fig.~\ref{fig:phasediagram}(a).

There is no external magnetic field in the calculation, but the Maxwell equations for the vector potential
should be solved self-consistently with the order parameter when the spontaneous magnetic field appears
in the ${\cal T}$-symmetry broken state.
For small grains $D\ll\lambda_0$, where $\lambda_0$ is the London penetration depth at $T=0$,
the influence of this self-consistency is very small and can safely be neglected.
We have also checked that for grains of sizes $D\gtrsim\lambda_0$, the vortex phase is unaffected. The vortices
formed spontaneously have a radius of a few coherence lengths, much smaller than the penetration depth
which for instance for YBCO is of order $\lambda_0\sim 100\xi_0$. The staggered ordering of these vortices means that
there is always only a very small effect of including the electrodynamics self-consistently in the calculation.
We only expect corrections for weak type-II superconductors where screening effects are more efficient 
(possibly other unconventional superconductors than the cuprates).

In Fig.~\ref{fig:sprops}(a) we plot the magnitude of the surface currents close to the interface.
The typical distance between current-nodes (neighboring vortex cores) on the surface is of the order of $5 \xi_0$. The
magnetic-field pattern is shown in \ref{fig:sprops}(b). The maximum magnetic field in the center of each fractional vortex 
is $1.5 \times 10^{-5} \Phi_0/\xi_0^2$ and we approximate the magnetic flux per vortex to $\sim10^{-5} \Phi_0$.  
In Fig.~\ref{fig:sprops}(c) we plot the low-energy part of the local DOS along a cut at $y\approx\xi_0$
marked with the black dashed line in (a).
The spectrum is split into two branches that meander between positive and negative energies. These two branches consists
of Andreev states carrying opposite currents along the surface, as also shown in Fig.~\ref{fig:DOS}(b)-(c).
In the symmetry broken phase, the Andreev surface states are pushed away from zero energy by the Doppler shifts.
The Andreev states at zero energy (for $T>T_{cs}$) are associated with a surface order parameter suppressed to zero. 
As ${\cal T}$-symmetry is broken, the $d$-wave order parameter partially heals, as shown in Fig.~\ref{fig:sprops}(d). At the same time,
its magnitude is oscillating along the surface, following the meandering of the Andreev states in the vortex pattern.

The low-energy states, positioned at $\varepsilon_A$, come at a cost in free energy.
If the bound states are located near the Fermi energy, $\varepsilon_A\to 0$,
it is energetically favorable to shift the states so that $\varepsilon_A<0$.
If the low-energy states have a substantial spectral weight, additional symmetries can be broken spontaneously,
such that the bound states shifts are guaranteed by the order parameter of the low-temperature phase.
There are competing mechanisms for this to happen, for instance spontaneous generation of orbital currents
or ferromagnetism. For the case of orbital currents, which is the focus of this work,  the order parameter field will 
acquire a phase gradient which is equivalent to a current that will Doppler shift the states to lower energies.
This is analogous to the Fulde-Ferrel-Larkin-Ovchinnikov (FFLO) instability,
where a superconductor generates a spatial modulation in response to
a Pauli coupling to a high magnetic field \cite{FuldeFerrel,LarkinOvchinnikov}.
For unconventional superconductors it was first shown using Ginzburg-Landau theory that a bulk superconductor,
in this case the heavy fermion compound UPt$_3$, may lower its free energy by generating
a spontaneous orbital current \cite{Palumbo1990}.
In the case of the $d$-wave high-temperature superconductors it was shown that at junctions between two crystals
there would be a spontaneous phase jump over the junction, both in the tunnelling limit\cite{Sigrist1995} and at general
transparencies \cite{Fogelstrom1998}, to minimize the junction free energy.
This effect leads to phase frustration in tri-crystal junctions that can be measured as a
half-quantum vortex located at the center of the tri-crystal \cite{Tsuei1994}.
At single surfaces in $d$-wave superconductors the low-lying states lay exactly at the Fermi level.\cite{Hu1994}
These states may be shifted away from the Fermi level by a Doppler shift \cite{Fogelstrom1997}.
At temperatures below $T\lesssim (\xi_0/\lambda_0)T_{\rm c}$ this may occur by spontaneous generation
of a finite superfluid momentum, or equivalently, phase gradient in
the order parameter \cite{Higashitani1997,Barash2000,Lofwander2000}.
This effect may then be detected as an anomaly in the low-temperature penetration depth.
But as the ratio $\xi_0/\lambda_0\lesssim0.01$ for the cuprates it may be hard to detect.
For thin films of $d$-wave superconductors \cite{anton}, it has been shown that transverse confinement can lead to
formation of a state with broken ${\cal T}$-symmetry at elevated temperatures larger than $(\xi_0/\lambda_0)T_c$
found for a single surface discussed above.
The transition temperature $T_{cs}$ for thin films scales approximately as the inverse of the film thickness before it becomes
unstable and the state with broken translational invariance in the form of a necklace of fractional vortices becomes favourable,
as seen in Fig.~\ref{fig:phasediagram}(b). We note that since the order parameter is inhomogeneous,
odd-frequency pairing correlations will be associated with the vortex necklace, as in any inhomogeneous superconducting
state \cite{Eschrig2007,Higashitani2014,Suzuki2014}.

Let us next discuss the consequences of the above described low-temperature phase for experiments.
Direct measurement of the induced magnetic fields associated with the vortex pattern might at first appear challenging
because of the small size of the vortices. However, recently, a scanning superconducting quantum interference device
was fabricated with a loop diameter of only $46\, {\rm nm}$ \cite{nanoSQUID}.
Vortices separated by about $120\, {\rm nm}$ were possible to detect.
The vortices in the symmetry broken state above are in our calculation separated by about $5\xi_0\sim 10\, {\rm nm}$,
within reach with state-of-the-art scanning nano-SQUIDS in the near future.

As a consequence of the Doppler shifts in the symmetry broken state, the local DOS at any specific
surface location contains a split low-energy peak, see Fig.~\ref{fig:sprops}(c). This split is a fingerprint
of broken ${\cal T}$-symmetry, either involving a subdominant component of the order parameter,
or not (as in the above calculation)  
But the split peak has been very difficult to measure in tunneling experiments and the possibility of a
low-temperature ${\cal T}$-symmetry broken phase in the cuprates remains controversial.
Instead of a split peak, the low-energy tunnelling spectrum typically features a broad peak also at low temperature. 
Our calculation offers a possible explanation for why the observation of broken $\cal T$-symmetry
has been so difficult. The tunnel junctions typically have rather large areas which
means that the local DOS should be integrated over a large surface area, larger than the vortex pattern.
The resulting area-integrated DOS display a very broad peak, see Fig.~\ref{fig:DOS}(a),
instead of the characteristic split Andreev peaks usually associated with the ${\cal T}$-symmetry broken state.
To observe the split peak, a junction diameter of the order of a few nanometers is needed.

We have also performed calculations with a subdominant $s$-wave pairing channel,
which in the symmetry broken state leads to a $d+is$ superconducting state
with the $s$-wave component appearing near the surface. The vortex necklaces are formed also
in this case. In addition we find that the broad DOS around $\varepsilon=0$ splits into two peaks as
is shown in Fig.~\ref{fig:DOS}(a). For small grains, when the size of the grain becomes comparable with 
the size of an individual vortex, we recover the results of those in Ref.~\cite{BlackSchaffer2013}. 

In summary, we have described a low-temperature phase of $d$-wave superconducting grains
with simultaneously broken ${\cal T}$-symmetry and broken translational symmetry along the grain surface.
In this phase a vortex pattern is formed, within which neighboring vortices have opposite current circulation.
The vortices have sizes of the order of a few coherence lengths, and can be measured directly
with recently developed scanning nano-SQUIDs \cite{nanoSQUID}.

We thank the Swedish research council for financial support, and Anton Vorontsov for valuable discussions.
\\
\\
{\bf Methods:}
We use the quasiclassical theory of superconductivity (see for instance \cite{SereneRainer})
to self-consistently solve for the superconducting state of mesoscopic grains.
The central object of the theory is the quasiclassical Green's function $\hat g(\pfhv,\vR;z)$,
that depends on momentum on the Fermi surface $\pfhv$, spatial coordinate $\vR$, and energy,
either real energy for the retarded/advanced Green's functions $z=\varepsilon\pm i0^+$ (infinitesimal imaginary part $i0^+$),
or Matsubara frequency $z=i\varepsilon_n=i\pi T(2n+1)$ in the Matsubara technique ($T$ is the temperature and $n$ an integer).
We utilize the so-called Riccati formulation \cite{nagato1993,schopohl1995}
and parametrize the Green's function by two coherence functions $\gamma(\pfhv,\vR;z)$ and $\tilde \gamma(\pfhv,\vR;z)$ as
\begin{equation}
\hat g^{R}=- i \pi \hat N 
\left(\begin{array}{cc}
1+ \gamma \tilde \gamma & 2 \gamma \\
-2 \tilde \gamma &-1 -\tilde \gamma \gamma \end{array}\right)
=\left(\begin{array}{cc}g&f\\ 
\tilde f&\tilde g\end{array}\right),
\label{qcgmatrix}
\end{equation}
with $\hat N={\rm diag}[(1- \gamma \tilde \gamma)^{-1},(1- \tilde \gamma \gamma)^{-1}]$.
To keep the notation compact, we suppress the dependences of all functions on $\pfhv$, $\vR$, and $z$.
Quantities with and without "tilde" are related by the symmetry $\tilde \alpha(\pfhv,\vR;z)=\alpha(-\pfhv,\vR;-z^*)^*$.
The coherence function obey a Riccati equation
\begin{equation} 
i\vvF\!\cdot\!\nabla \gamma  +2 z  \gamma = -\gamma \tDelta  \gamma  -\Delta,
\label{gamman}
\end{equation}
%
%
which is solved by integration along straight lines,
or trajectories
$\vs(x)=\vs_0 + x\, \vvF/\vert\vvF\vert$,
as described in e.g. Ref.~\cite{sauls2009}.
At the boundaries we assume specular reflection. The system of equations is closed by the gap equation
\begin{equation}
\Delta_d(\vR)\!=\lambda_{d} T\sum_{|\varepsilon_n|\le\varepsilon_{c}} \langle {\cal{Y}}^*_{d}(\pfhv)f(\pfhv,\vR;\varepsilon_n)
\rangle_{\pfhv},
\label{OPs}
\end{equation}
where $\langle \cdots \rangle_{\pfhv}=\int\frac{d\phi_p}{2\pi}$ denotes the average over the momentum direction
on the Fermi surface, with the angle $\phi_p$ giving the angle the momentum $\pfhv$ makes to the crystal $a$-axis.
The conventional basis function ${\cal{Y}}_{d}(\pfhv)=\sqrt{2}\cos(2\phi_p)$ for a $d$-wave superconducting state is used. 
The paring interaction $\lambda_d$ and the cut-off frequency $\varepsilon_c$ are eliminated in favour of the transition temperature
$T_c$, as $\lambda_{d}^{-1}=\ln(T/T_c)+\sum_{n\ge 0}(n+1/2)^{-1}$. The above equations are solved numerically 
and iteratively on a graphics card until self-consistency has been achieved.

The free energy functional has been written down by Eilenberger \cite{Eilenberger1968},
\begin{align}
&\delta \Omega[T]=\int d\vR \Bigg\lbrace
\frac{\vB(\vR)^2}{8\pi}
+|\Delta(\vR)|^2 N_F\, {\rm ln} \frac{T}{T_c}\nonumber\\
&+2\pi N_F k_B T\sum^{\infty}_{n=0} \bigg\lbrack\frac{|\Delta(\vR)|^2}{\varepsilon_n}+i
\langle {\cal{I}}(\pfhv,\vR;\varepsilon_n)\rangle_{\pfhv}  \bigg\rbrack
\Bigg\rbrace
\end{align}

where ${\cal{I}}=\Delta^*(\pfhv,\vR)\gamma(\pfhv,\vR;\varepsilon_n)-\Delta(\pfhv,\vR)\tilde\gamma(\pfhv,\vR;\varepsilon_n)$, 
$N_F$ is the normal-state density of states at the Fermi level,
and $\vB(\vR)$ is the generated magnetic field.

\end{document}